\documentclass[namedreferences]{SolarPhysics}
\usepackage[optionalrh]{spr-sola-addons} 

\usepackage[dvips]{graphicx}
\usepackage{exscale}
\usepackage{amsfonts}
\usepackage{amssymb}
\usepackage{url}             

\DeclareMathAlphabet{\mathbfl}{OML}{cmm}{b}{it} 
\newcommand{\vect}[1]{\ensuremath{\mathbfl{#1}}} 
\newcommand{\lvecc}[2]{\ensuremath{(#1,#2)}}
\newcommand{\text}[1]{\ensuremath{\mbox{\rm #1}}}
\newcommand{\abs}[1]{\ensuremath{|#1|}}
\newcommand{\grd}[1]{\ensuremath{\mathbf{\nabla}} #1}
\newcommand{\vectg}[1]{\ensuremath{\mathbf{#1}}} 
\newcommand{\half}{\ensuremath{{\textstyle \!\frac{1}{2}}}}

\begin{document}
\begin{article}
\begin{opening}

\title{Segmentation of Loops from Coronal EUV Images}

\author{B.~{Inhester}$^{1}$,
        L.~{Feng}$^{1,2}$, and
        T.~{Wiegelmann}$^{1}$}

\runningauthor{Inhester et al.}
\runningtitle{Segmentation of loops from coronal EUV images}

\institute{$^{1}$ Max-Planck-Institut f\"ur Sonnensystemforschung, Max-Planck-Str.2,
                  37191 Katlenburg-Lindau, Germany
                  \email{Inhester@mps.mpg.de}\\
           $^{2}$ Purple Mountain Observatory, Chinese Academy of Sciences, Nanjing, China\\}

\date{Received: 26 April 2007, ~accepted: July 31 2007\\
      doi: 10.1007/s11207-007-9027-1}

\begin{abstract}
 We present a procedure which extracts bright loop features from solar EUV
images. In terms of image intensities, these features are elongated
ridge-like intensity maxima. To discriminate the maxima, we need
information about the spatial derivatives of the image intensity.
Commonly, the derivative estimates are strongly affected by image
noise. We therefore use a regularized estimation of the derivative
which is then used to interpolate a discrete vector field of ridge
points ``ridgels'' which are positioned on the ridge center and have the
intrinsic orientation of the local ridge direction. A scheme is
proposed to connect ridgels to smooth, spline-represented curves which
fit the observed loops. Finally, a half-automated user interface
allows one to merge or split, eliminate or select loop fits obtained form
the above procedure.
In this paper we apply our tool to one of the first EUV images
observed by the SECCHI instrument onboard the recently launched
STEREO spacecraft. We compare the extracted loops with projected field
lines computed from almost-simultaneously-taken magnetograms measured
by the SOHO/MDI Doppler imager. The field lines were calculated
using a linear force-free field model.
This comparison
allows one to verify faint and spurious loop connections produced by our
segmentation tool and it also helps to prove the quality of the
magnetic-field model where well-identified loop structures comply with
field-line projections. We also discuss further potential applications
of our tool such as loop oscillations and stereoscopy.
\end{abstract}
\end{opening}

\section{Introduction}

Solar EUV images offer a wealth of information about the structure of
the solar chromosphere, transition region, and corona. Moreover, these
structures are in continuous motion so that the information collected
by EUV images of the Sun is enormous. For many purposes this
information must be reduced.
A standard task for many applications, {\it e.g.}, for the comparison with
projected field lines computed from a coronal magnetic-field model or
for tie-point stereoscopic reconstruction, requires the extraction ofthe
shape of bright loops from these images.

Solar physics shares this task of ridge detection with many other
disciplines in physics and also in other areas of research. A wealth
of different approaches for the detection and segmentation of
ridges has been proposed ranging from multiscale filtering
\cite{Koller:etal:1995,Lindeberg:1998} and curvelet and ridgelet transforms
\cite{Starck:etal:2003} to snake and watershed algorithms
\cite{nguyen:etal:2000} and combining detected ridge points by tensor
voting \cite{Medioni:etal:2000}. These general methods however always need
to be modified and optimized for specific applications. Much work
in this field has been motivated by medical imaging
({\it e.g.}, \opencite{Jang:Hong:2002}; \opencite{Dimas:etal:2002}) and
also by the application in more technical fields like fingerprint
classification \cite{Zhang:Yan:2004} and the detection of roads in
areal photography \cite{Steger:1998}.

For the automated segmentation of loops, a first step was made by
Strous (2002, unpublished) who proposed a procedure to detect pixels
in the vicinity of loops. This approach was further extended by
\inlinecite{Lee:etal:2006} by means of a connection scheme which makes use
of a local solar-surface magnetic-field estimate to obtain
a prefereable connection orientation. The procedure then leads
to spline curves as approximations for the loop shapes in the image. 
The method gave quite promising results for artificial and also
for observed trace EUV images.

The program presented here can be considered an extension of the work
by \inlinecite{Lee:etal:2006}. The improvements which we propose are
to replace Strous' ridge-point detection scheme by a modified multiscale
approach of \inlinecite{Lindeberg:1998} which automatically adjusts to
varying loop thicknesses and returns also an estimate of the reliability
of the ridge point location and orientation.
When connecting the ridge points, we would like not to use any magnetic
field information as this prejudices a later comparison of the extracted
loops with field lines computed from an extrapolation of the surface
magnetic field. As we consider this comparison a validity test for the
underlying field extrapolation, it would be desirable to derive the
loop shapes independently. Our connectivity method is therefore
based only on geometrical principles which combines the orientation
of the loop at the ridge point with the cocircularity constraint
proposed by \inlinecite{Parent:Zucker:1989}.

The procedure is performed in three steps, each of which could be
considered a module of its own and performs a very specific task. In
the following sections we explain these individual steps in some
detail. In the successive section we apply the scheme to one of the
first images observed by the SECCHI instruments onboard the
recently-launched STEREO spacecraft \cite{Howard:etal:2007} in order to
demonstrate the capability of our tool. Our procedure offers
alternative subschemes and adaptive parameters to be adjusted to the
contrast and noise level of the image to be dealt with. We discuss how
the result depends on the choice of some of these parameters. In the final
section we discuss potential applications of our tool.

\section{Method}

Our approach consists of three modular steps, each of which is
described in one of the following subsections. The first is to find
points which presumably are located on the loop axis. At these
positions, we also estimate the orientation of the loop for these
estimates. Each item with this set of information is called a ridgel.
The next step is to establish probable neighbourhood relations between
them which yields chains of ridgels. Finally, each chain is fitted by
a smoothing spline which approximates the loop which gave rise to the
ridgels.

\subsection{Ridgel Location and Orientation}

In terms of image intensities, loop structures are elongated ridge-like
intensity maxima. To discriminate the maxima, we need information about
the spatial derivatives of the image intensity. Commonly, these
derivatives are strongly affected by image noise. In fact, numerical
differentiation of data is an ill-posed problem and calls for proper
regularization.

We denote by $\vect{i}$ $\in$ $\mathbb{I}^2$ the integer coordinate
values of the pixel centres in the image and by $\vect{x}$ $\in$
$\mathbb{R}^2$ the 2D continuous image coordinates with $\vect{x}$ =
$\vect{i}$ at the pixel centres.
We further assume that the observed image intensity $I(\vect{i})$ varies
sufficiently smoothly so that a Taylor expansion at the cell centres
is a good approximation to the true
intensity variation $I(\vect{x})$ in the neighbourhood of $\vect{i}$,
{\it i.e.},
\begin{equation}
  I(\vect{x}) \simeq \tilde{I}(\vect{x}) =  c + \vect{g}^T(\vect{x}-\vect{i})
    + (\vect{x}-\vect{i})^T\vect{H}(\vect{x}-\vect{i})
\label{eq:ITaylor}\end{equation}
Pixels close to a ridge in the image intensity can then be detected
on the basis of the local derivatives $\vect{g}$ and $\vect{H}$
(the factor 1/2 is absorbed in $\vect{H}$). We
achieve this by diagonalizing $\vect{H}$, {\it i.e.}, we determine the unitary
matrix $\vect{U}$ with
\begin{equation}
  \vect{U}^T\vect{H}\vect{U} = \mathrm{diag}(h_\perp,h_\parallel)
  \quad\text{where}\quad\vect{U}=\lvecc{\vect{u}_\perp}{\vect{u}_\parallel}
\end{equation}
where we assume that the eigenvector columns $\vect{u}_\perp$ and
$\vect{u}_\parallel$ of $\vect{U}$ associated to the eigenvalues $h_\perp$
and $h_\parallel$, respectively, are ordered so that $h_\perp$ $\le$
$h_\parallel$.

We have implemented two ways to estimate the Taylor coefficients.
The first is a local fit of (\ref{eq:ITaylor}) to the image within
a $(2m+1)\times (2m+1)$ pixel box centered around each pixel $\vect{i}$:
\begin{equation}
    (c,\vect{g},\vect{H})(\vect{i}) = \mathrm{argmin}
    \!\!\!\!\!\!\!\!\!\!\!\!\!\!\!\!\!
    \sum\limits_{\vect{j}-\vect{i}\in[-m,m]\times[-m,m]}
    \!\!\!\!\!\!\!\!\!\!\!\!\!\!\!\!\!
    w(\vect{i}-\vect{j})\;\big(\tilde{I}(\vect{j})-I(\vect{j})\big)^2
\label{eq:TaylorFit}\end{equation}
We use different weight functions $w$ with their support limited to the 
box size such as triangle, cosine or cosine$^2$ tapers. 

The second  method commonly used is to calculate the Taylor coefficients
(\ref{eq:ITaylor}) not from the original but from a filtered image
\begin{equation}
 \bar{I}(\vect{x}) = \sum_\vect{j} w_d(\vect{x}-\vect{j}) I(\vect{j})
\label{eq:TaylorFilter}\end{equation}
As window function $w_d$ we use a normalised Gaussian of width $d$.
The Taylor coefficients can now be explicitly derived by differentiation
of $\bar{I}$, which however acts on the window function $w_d$ instead on
the image data. We therefore effectively use a filter kernel for each Taylor
coefficient which relates to the respective derivatives of the window function
$w_d$.

One advantage of the latter method over the local fit described above is
that the window width $d$ can be chosen from $\mathbb{R}_+$ while the
window size for the fit procedure must be an odd integer $2m+1$.
Both of the above methods regularize the Taylor coefficient estimate by the
finite size of their window function. In fact, the window size could be
considered as a regularization parameter.

A common problem of regularized inversions is the proper choice of the
regularization parameter. \inlinecite{Lindeberg:1998} has devised
a scheme for how this parameter can be optimally chosen. Our third method
is a slightly modified implementation of his automated scale
selection procedure. The idea is to apply method 2 above for each
pixel repeatedly with increasing scales $d$ and thereby obtain an
approximation of the ridge's second derivative eigenvalues $h_\perp$ and
$h_\parallel$, each as a function of the scale $d$.

Since the $h_\perp$ and $h_\parallel$ are the principal second order
derivatives of the image after being filtered with $w_d$, they depend
on the width $d$ of the filter window roughly in the following way.
As long as $d$ is much smaller than the intrinsic width of the ridge,
$d_\perp$ = $\abs{(\vect{u}_\perp^T\grd{})^2{\log I}}^{-1/2}$, the
value in $h_\perp$ will be a (noisy) estimate of the true principal
second derivative $(\vect{u}_\perp^T\grd{})^2 I$ of the image,
independent of $d$. Hence, $h_\perp \propto -I_\mathrm{max}/d_\perp^2$
for $d^2\ll d_\perp^2$. To reduce the noise and enhance the significance
of the estimate, we would however like to choose $d$ as large as possible.
For $d^{2}$ $\gg$ $d_\perp^2$, the result obtained for $h_\perp$ will
reflect the shape of the window rather that of the width of the ridge,
$h_\perp \propto -\bar{I}_\mathrm{max}/d^2$ = $-I_\mathrm{max} d_\perp/d^3$ 
for $d^2\gg d_\perp^2$. Roughly, details near $d$ $\sim$ $d_\perp$ depend
on the exact shape of the ridge, we have
\begin{equation}
h_\perp(d) \sim \frac{-d_\perp}{\big(d_\perp^2+d^2\big)^{3/2}}
\end{equation}
For each pixel we consider in addition a quality function
\begin{equation}
  q(d)= d^{\gamma}(\abs{h_\perp}-\abs{h_\parallel})\;,\quad \gamma\in(0,3)
\label{eq:qmx}\end{equation}
which will vary as $d^{\gamma}$ for small $d$ and decrease asymptotically to zero
for $d\gg d_\perp$ as $d^{\gamma-3}$. In between, $q(d)$ will reach a maximum
approximately where the window width $d$ matches the local width $d_\perp$ of
the ridge (which is smaller than the scale along the ridge).
The choice of the right width $d$ has now been replaced by a choice for the
exponent $\gamma$. The result, however, is much less sensitive to variations
in $\gamma$ than to variations in $d$. Smaller values of $\gamma$ shift the
maximum of $q$ only slightly to smaller values of $d$ and hence tend to
favour more narrow loop structures.
While $\gamma$ is a constant for the whole image in this automated scale
selection scheme, the window width $d$ is chosen individually for every
pixel from the respective maximum of the quality factor $q$.

\begin{figure} 
  \includegraphics[width=6cm]{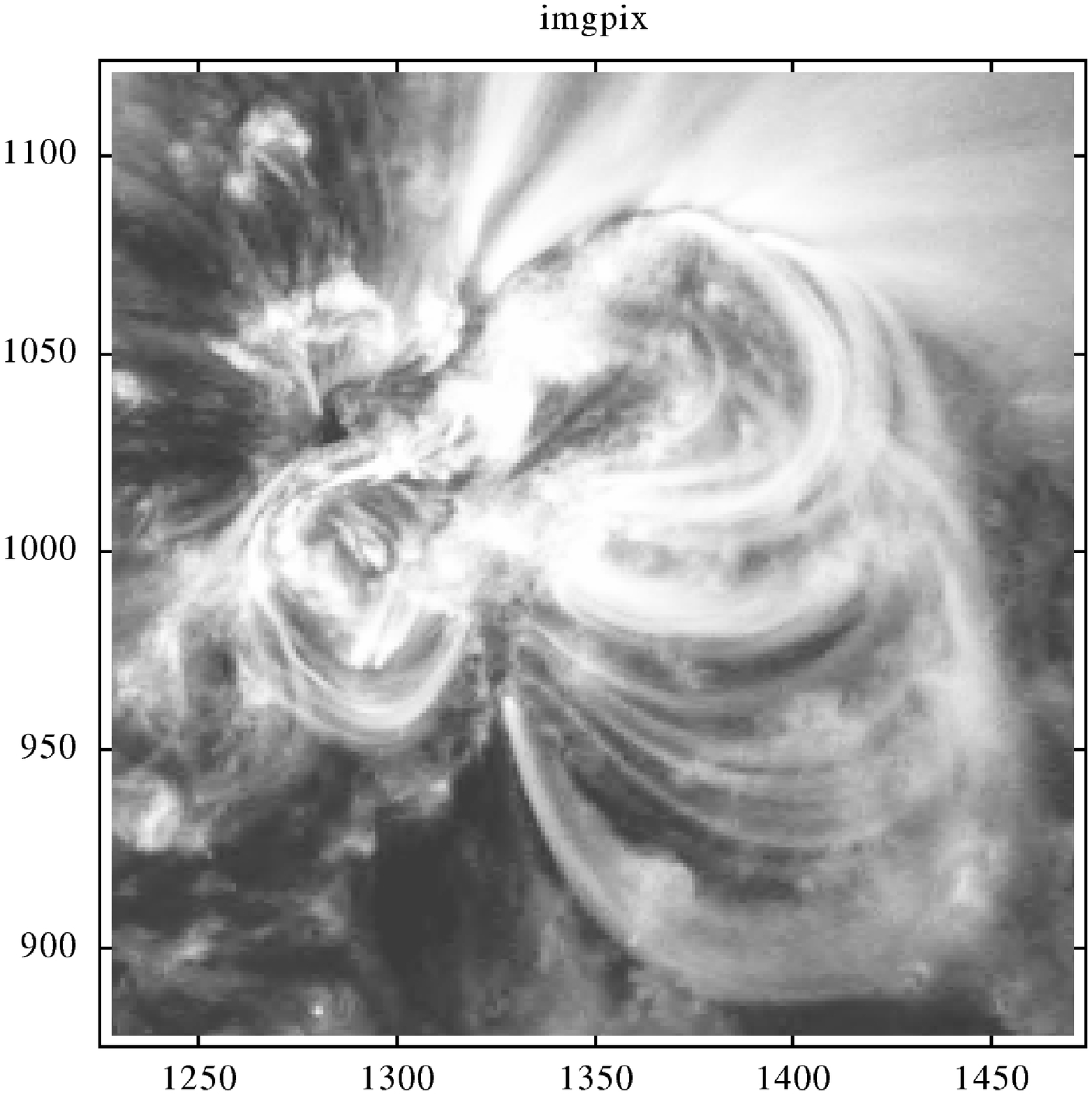}
  \includegraphics[width=6cm]{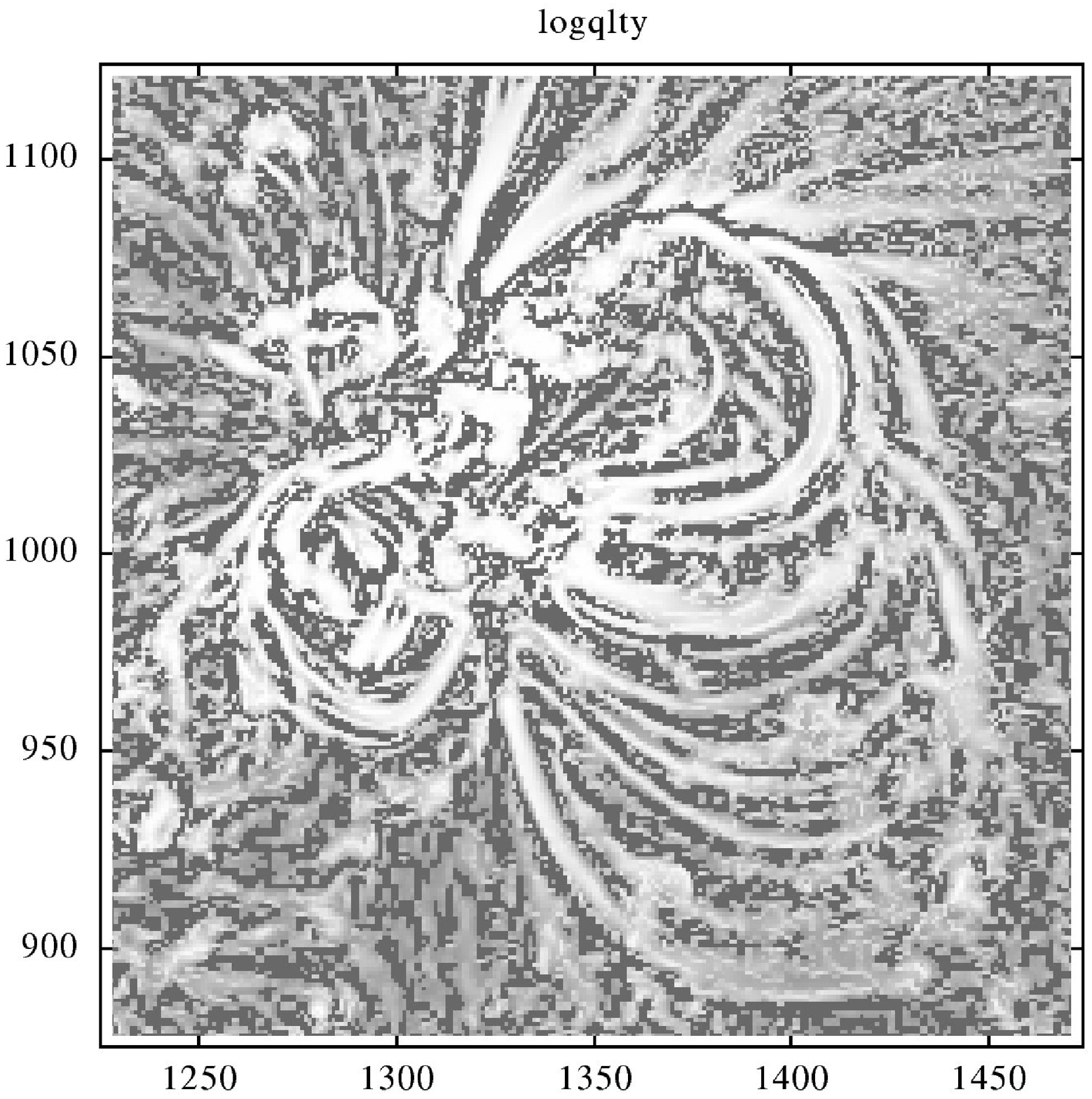}
\caption{Original image of active region NOAA 10930 observed by STEREO/SECCHI
(left) and the corresponding image of the quality factor $q$ (\ref{eq:qmx})
obained from the third ridgel determination method by automated scale
selection (right). This image was taken at $\lambda$ = 171 {\AA} on 12 December 2006 at
20:43 UT and was not processed by the SECCHI\_prep routine.}\label{fig:qmx}
\end{figure}

\begin{figure}
 \hspace*{\fill}
  \includegraphics[bb=0 0 440 334,clip,width=6cm,height=4.7cm]{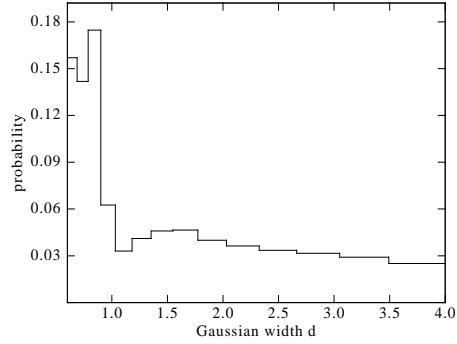}
\hspace*{\fill}
\caption{Distribution of widths $d$ of the window function $w_d$ for which
  (\ref{eq:qmx}) was found to reach a maximum when applied to the data in
  image (\ref{fig:qmx}).
  The maximum was independently determined for each image pixel for which
  $h_\perp<\abs{h_\parallel}$.}\label{fig:qprob}
\end{figure}

In Figure~\ref{fig:qmx} we show as an example a $\lambda$ = 171 {\AA} image
of active region NOAA 10930 observed by STEREO/SECCHI on 12 December 2006 at
20:43 UT and the corresponding image of $q$ obtained with $\gamma=0.75$
and window sizes $d$ in the range of 0.6 to 4 pixels. Clearly, the
$q$ factor hass a maximum in the vicinity of the loops. The distribution of
the scales $d$ for which the maximum $q$ was found for each pixel is shown
in Figure~\ref{fig:qprob}. About 1/3 of the pixels had optimal widths $<$ one pixel,
many of which originate from local longated noise and moss features of the
image. The EUV moss is an amorphous emission which originates in the
upper transition region \cite{TEBerger:etal:1999} and are not
associated with loops.
For proper loop structures the optimum width found was about
1.5 pixels with, however, a widely spread distribution.

In the case that $\vect{i}$ is located exactly on a ridge, $\vect{u}_\perp$
is the direction across and $\vect{u}_\parallel$ the direction along
the ridge and $h_\perp$ and $h_\parallel$ are the associated second
derivatives of the image intensity in the respective direction. A
positive ridge is identified from the Taylor coefficients by means
of the following conditions \cite{Lindeberg:1998}
\begin{eqnarray}
  \vect{u}_\perp^T\vectg{\nabla}I=\vect{u}_\perp^T\vect{g}=0
  &&\text{a vanishing gradient across the ridge}
  \label{eq:ridgecrit1}\\
  (\vect{u}_\perp^T\vectg{\nabla})^2I=h_\perp<0
  &&\text{a negative second order derivative}
  \label{eq:ridgecrit2}\\[-0.5ex]
  &&\text{across the ridge}
  \nonumber\\
  \abs{(\vect{u}_\perp^T\vectg{\nabla})^2I}>
  \abs{(\vect{u}_\parallel^T\vectg{\nabla})^2I}
  &&\text{a second order derivative magnitude}\\[-0.5ex]
  \label{eq:ridgecrit3}
  \text{or}\quad\abs{h_\perp}>\abs{h_\parallel}\hspace*{2.9em}
  &&\text{across the ridge larger than along}
  \nonumber
\end{eqnarray}
The latter two inequalities are assumed to also hold in the near
neighbourhood of the ridge and are used to indicate whether the
pixel centre is close to a ridge.

In the vicinity of the ridge, along a line
$\vect{x}$ = $\vect{i}+\vect{u}_\perp t$, the image
intensity (\ref{eq:ITaylor}) then varies as
\begin{equation}
  I(t) \simeq c + \vect{u}_\perp^T\vect{g} t
                + \vect{u}_\perp^T\vect{u}_\perp h_\perp t^2
\end{equation}
According to the first ridge criterion (\ref{eq:ridgecrit1}), the precise
ridge position is where $I(t)$ has its maximum.
Hence the distance to the ridge is
\begin{equation}
  t_\mathrm{max}=-\frac{\vect{u}_\perp^T\vect{g}}{2h_\perp\vect{u}_\perp^T\vect{u}_\perp}
\end{equation}
and a tangent section to the actual ridge curve closest to $\vect{i}$ is
\begin{equation}
  \vect{r}(s)=\vect{i}-\frac{\vect{u}_\perp\vect{u}_\perp^T\vect{g}}
                            {2h_\perp\vect{u}_\perp^T\vect{u}_\perp}
          + s\vect{u}_\parallel \quad\text{for}\quad s\in\mathbb{R}
\label{eq:ridgept1}\end{equation}
Note that $\vect{u}_\perp^T\vect{u}_\perp$ = 1 for a unitary $\vect{U}$.

\begin{figure} 
  \includegraphics[bb=0 0 245 307,clip,width=6cm]{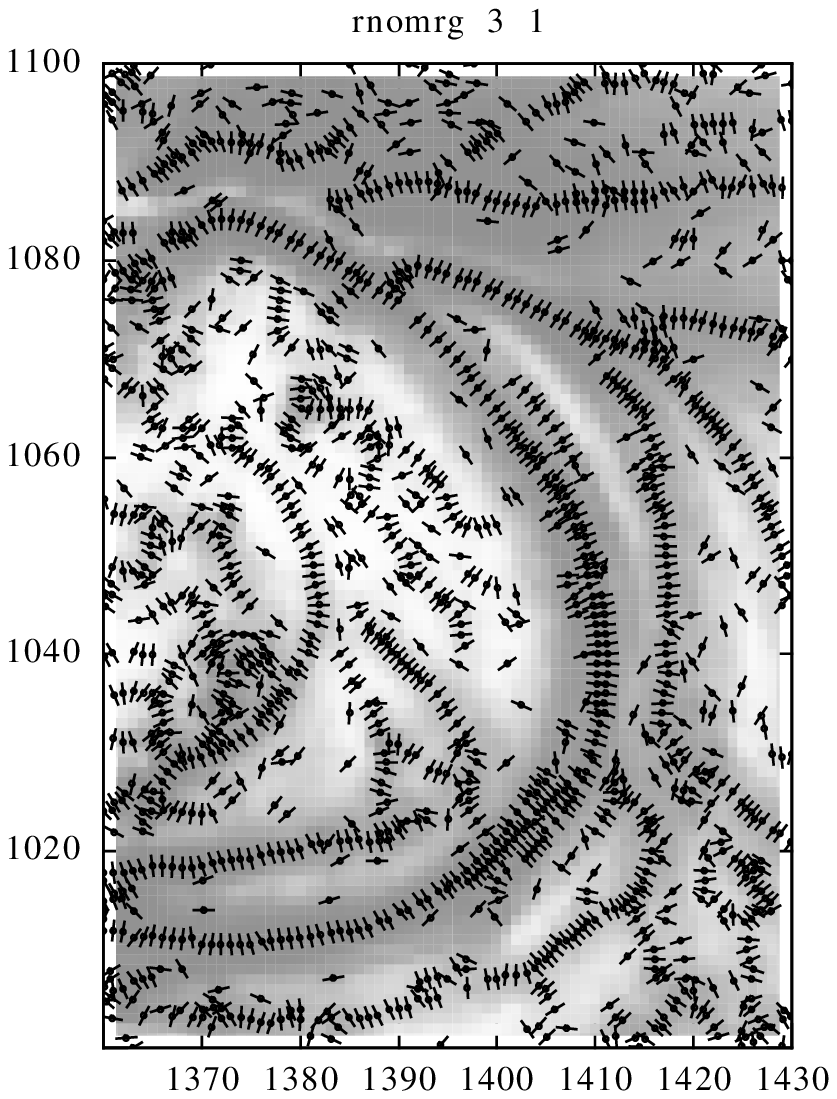}
  \includegraphics[bb=0 0 245 307,clip,width=6cm]{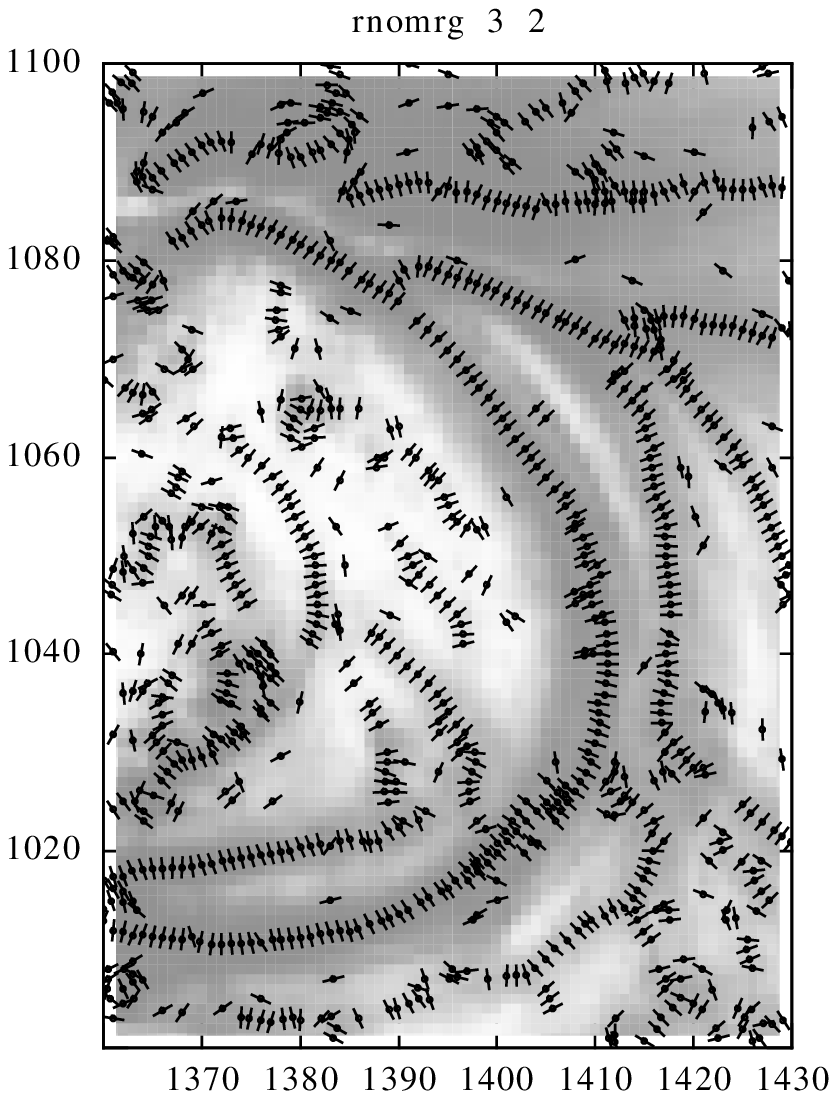}
\caption{Comparison of the resulting ridgels from interpolation method
 (\ref{eq:ridgept1}, left) and (\ref{eq:ridgept2}, right).
 The images show an enlarged portion of the original data in
 Figure~\ref{fig:qmx}. The short sticks denote the local orientation
 of $\vect{u}_\perp$ on the loop trace.)}\label{fig:ridgels}
\end{figure}

We have implemented two methods for the interpolation of the ridge
position from the Taylor coefficients calculated at the pixel centres.
One is the interpolation of the ridge centre with the help of
(\ref{eq:ridgept1}). The second method interpolates the zeros of
$\vect{u}_\perp^T\vect{g}$ in between neighbouring pixel centres
$\vect{i}$ and $\vect{j}$ if its sign changes. 
Hence the alternative realization of (\ref{eq:ridgecrit1}) is
\[
 \text{if}\quad\abs{c}=\abs{\vect{u}_\perp^T(\vect{i})\vect{u}_\perp(\vect{j})}
                      >c_\mathrm{min}
\]\vspace*{-4ex}\[
 \text{and}\quad \mathrm{sign}(c)(\vect{u}_\perp^T\vect{g})(\vect{j})\;
                                (\vect{u}_\perp^T\vect{g})(\vect{i})<0 
  \quad\text{then}
\]\begin{equation}
  \vect{r}=\vect{i} + t(\vect{j}-\vect{i})
  \label{eq:ridgept2}
\end{equation}\[
  \text{where}\quad
           t=\frac{(\vect{g}^T\vect{u}_\perp)(\vect{i})}
                   {(\vect{g}^T\vect{u}_\perp)(\vect{i})
   -\mathrm{sign}(c)(\vect{g}^T\vect{u}_\perp)(\vect{j})}
\]
The first condition insures that $\vect{u}_\perp(\vect{i})$ and
$\vect{u}_\perp(\vect{j})$ are sufficiently parallel or antiparallel.
Note that the orientation of $\vect{u}_\perp$ of neighbouring pixels
may be parallel or antiparallel because an eigenvector $\vect{u}_\perp$
has no uniquely defined sign.

In general, the interpolation according to (\ref{eq:ridgept1}) yields
fewer ridge points along a loop but they have a fairly constant relative
distance.
With the second method (\ref{eq:ridgept2}), the ridge points can only be
found at the intersections of the ridge with the grid lines connecting
the pixel centres. For ridges directed obliquely to the grid, the
distances between neighbouring ridge points produced may vary by some
amount.
Another disadvantage of the second method is that it cannot detect
faint ridges which are just about one pixel wide. It needs at least
two detected neighbouring pixels in the direction across the ridge to
properly interpolate the precise ridge position.
The advantage of the second method is that it does not make use of the
second order derivative $h_\perp$ which unavoidably is more noisy than the
first order derivative $\vect{g}$. In Figure~\ref{fig:ridgels} we compare the
ridgels obtained wth the two interpolation methods for the same image. 

The final implementation of identifying ridge points in the image
comprises two steps:
First the Taylor coefficients (\ref{eq:ITaylor}) are determined for
every pixel and saved for those pixels which have an
intensity above a threshold value $I_\mathrm{min}$, for which the
ridge shape factor $(h_\perp^2-h_\parallel^2)/(h_\perp^2+h_\parallel^2)$
exceeds a threshold $s_\mathrm{min}$ in accordance with (\ref{eq:ridgecrit3})
and which also satisfy (\ref{eq:ridgecrit2}).
The second step is then to interpolate the precise sub-pixel ridge point
position from the derivatives at these pixel centres by either of the
above methodes. This interpolation complies with the third ridge criterion
(\ref{eq:ridgecrit1}).
The ridgel orientation $\vect{u}_\perp$ are also interpolated from the
cell centres to the ridgel position. The information retained for
every ridge point $n$ in the end consists of its location
$\vect{r}_n$, and the ridge normal orientation $\vect{u}_{\perp,n}$
defined modulo $\pi$.

\subsection{Ridgel Connection to Chains}

The connection algorithm we apply to the ridgels to form chains
of ridgels is based on the cocircularity condition of
\inlinecite{Parent:Zucker:1989}.

\begin{figure}
 \hspace*{\fill}
  \includegraphics[bb=19 10 270 260,clip,height=6cm]{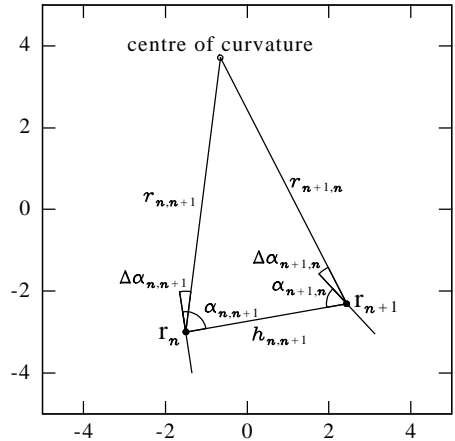}
 \hspace*{\fill}
\caption{\label{fig:cocirc} Illustration of angles and distances
  of a connection element between two ridgels according to the
  cocircularity condition of Parent and Zucker (1989).
  The ridgel positions are indicated by a small dot,
  the ridge normal orientation by the line centred at
  the ridgel position. The axis units are image pixels.}
\end{figure}

For two ridgels at $\vect{r}_n$ and $\vect{r}_{n+1}$ a virtual
centre of curvature can be defined which forms an isoceles triangle
with the ridgels as shown in Figure~\ref{fig:cocirc}.
One edge is formed by the connection between the two ridgels of
mutual distance $h_{n,n+1}$.
The two other triangle edges in this construction connect one of the two
ridgels with the centre of curvature which is chosen so that these two
symmetric edges of the isoceles triangle make angles
$\Delta\alpha_{n,n+1}$ and $\Delta\alpha_{n+1,n}$ as small as possible
with the respective ridgel orientation $\vect{u}_{\perp,n}$ and
$\vect{u}_{\perp,n+1}$, respectively.
It can be shown that
\begin{equation}
  \mathrm{min}~ (\Delta\alpha_{n,n+1}^2+\Delta\alpha_{n+1,n}^2)
\nonumber\end{equation}
requires equal magnitudes for the angles $\Delta\alpha_{n,n+1}$ and
$\Delta\alpha_{n+1,n}$.
The distance $r_{n,n+1}$ $=r_{n+1,n}$ is the local radius of curvature
and can be calculated from
\begin{equation}
  r_{n,n+1}=r_{n+1,n}=
  \frac{\half h_{n,n+1}}{\cos\big(\half(\alpha_{n,n+1}+\alpha_{n+1,n})\big)} 
\label{eq:curvrad}\end{equation}
where  $\alpha_{n,n+1}$ is the angle between $\vect{r}_{n+1}-\vect{r}_n$
and $\pm\vect{u}_{\perp,n}$, the sign being chosen so that
$\abs{\alpha_{n,n+1}}<\pi/2$.

With each connection between a pair of ridgels we associate a binding
energy which depends on the parameters derived above in the form:
\begin{equation}
  e_{n,n+1}= \big(\frac{\Delta\alpha_{n,n+1}}{\alpha_\mathrm{max}}\big)^2
            +\big(\frac{r_\mathrm{min}}{r_{n,n+1}}\big)^2
            +\big(\frac{h_{n,n+1}}{h_\mathrm{max}}\big)^2
            -3
\label{eq:cocirc}\end{equation}
Note that $\Delta\alpha_{n,n+1}^2$ = $\Delta\alpha_{n+1,n}^2$
according to the cocircularity construction and hence $e_{n,n+1}$ is symmetric
in its indices.
The three terms measure three different types of distortions and can
be looked upon as the energy of an elastic line element. The first
term measures the deviation of the ridgel orientation from strict
cocircularity, the second the bending of the line element, and the
third term its stretching.
The constants $\alpha_\mathrm{max}$, $r_\mathrm{min}$, and
$h_\mathrm{max}$ give us control on the relative weight of the three
terms. $r_\mathrm{min}$ is the smallest acceptable
radius of curvature and $h_\mathrm{max}$ is the largest acceptable distance
if we only accept connections with a negative value for the energy
(\ref{eq:cocirc}).

In practical applications, the energy (\ref{eq:cocirc}) is problematic
since it puts nearby ridgel pairs with small distances $h_{n,n+1}$ at a
severe disadvantage because small changes of their $\vect{u}_\perp$ 
easily reduces the radius of curvature $r_{n,n+1}$ below acceptable
values. We therefore allow for measurement errors in $\vect{r}$ and
$\vect{u}_\perp$ and the final energy considered is the minimum of
(\ref{eq:cocirc}) within these given error bounds. 

The final goal is to establish a whole set of connections between as
many ridgels as possible so that the the individual connections add up
to chains. Note that each ridgel has two ``sides'' defined by the two
half spaces which are separated by the ridgel orientation $\vect{u}_\perp$.
We only allow at most one connection per ridgel in each of these ``sides''.
This restriction avoids junctions in the chains that we
are going to generate. The sum of the binding energies
(\ref{eq:cocirc}) of all accepted connections ideally should
attain a global minimum in the sense that any alternative set of
connections which complies with the above restriction should yield a
larger energy sum.

We use the following approach to find a state which comes close to
this global minimum. The energy $e_{n,n+1}$ is calculated for each
ridgel pair less than $h_\mathrm{max}$ apart, and those connections
which have a negative binding energy are stored. These latter are the
only connections which we expect to contribute to the energy minimum.
Next we order the stored connections according to their energy and connect
the ridgels to chains starting from the lowest-energy connection.
Connections to one side of a ridgel which has already been occupied
by a lower energy connection before are simply discarded.

\subsection{Curve Fits to the Ridgel Chains}

\begin{figure}
 \hspace*{\fill}
  \includegraphics[bb=19 10 270 260,clip,width=6cm]{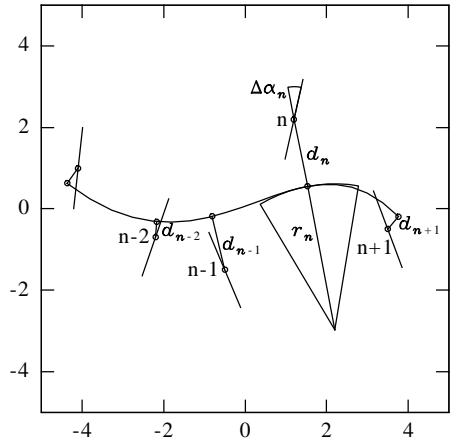}
 \hspace*{\fill}
\caption{\label{fig:curvefit}
 Sketch of the curve-fit parameters. The ridgels are represented by
 their location and the two-pixel-long bar of the rpc orientation.
 For each ridgel $i$, the proximity to the smooth-fit curve is expressed
 by their distance $d_i$ and the angle $\Delta\alpha_i$ between the
 distance direction of $\vect{d}_i$ to the curve and the rpc orientation.
 Another measure of the quality of the curve is the inverse
 radius of curvature $r$.}
\end{figure}

In this final section we calculate a smooth fit to the chains of ridgels
obtained above. The fit curve should level out small errors in the
position and orientation of individual ridgels.
We found from experiments that higher-order spline functions are far
too flexible for the curves we aim at.
We expect that magnetic-field lines in the corona do not rapidly vary
their curvature along their length and we assume this also holds for
their projections on EUV images.
We found that parametric polynomials of third or fifth degree
are sufficient for our purposes. Hence for each chain of ridgels
we seek polynomial coefficients $\vect{q}_n$ which generate a
two-dimensional curve
\begin{equation}
  \vect{p}(t) = \sum_{n=0}^5 \vect{q}_n t^n
  \quad\text{for}\quad t\in [-1,1]
\label{eq:polynomial}\end{equation}
which best approximates the chain of ridgels.
What we mean by ``best approximation'' will be defined more precisely
below. The relevant parameters of this approximation are sketched in
Figure~\ref{fig:curvefit}.

The polynomial coefficients $\vect{q}$ of a fit (\ref{eq:polynomial}) are
determined by minimising
\begin{equation}
     \sum_{i\in\mathrm{chain}} (\vect{d}_i^T\vect{d}_i)
                       + \mu (\vect{p''}^T\vect{p''})(t_i)
                       \label{eq:curvefit}
\end{equation}
\vspace*{-4ex}\[
    \text{where}\quad\vect{d}_i=\vect{r}_i - \vect{p}(t_i)
\]
with respect to $\vect{q}_n$ for a given $\mu$. Initially, we distribute
the curve parameters $t_i$ in the interval $[-1,1]$ such that the
differences $t_i-t_j$ of neighbouring ridgels are proportional to
the geometric distances $\abs{\vect{r}_i-\vect{r}_j}$.
The $\vect{p''}$ are the second-order derivatives of (\ref{eq:polynomial}).
Hence, the second term increases with increasing curvature of the fit while
a more strongly curved fit is required to reduce the distances $\vect{d}_i$
between $\vect{r}_i$ and the first order closest curve point $\vect{p}(t_i)$.  

The minimum coefficients $\vect{q}_n(\mu)$ can be found analytically
in a straight forward way. Whenever a new set of $\vect{q}_n(\mu)$ has
been calculated, the curve nodes $t_i$ are readjusted by
\begin{equation}
     t_i=\mathrm{argmin}_t (\vect{r}_i - \vect{p}(t))^2
\end{equation}
so that $\vect{p}(t_i)$ is always the point along the curve closest to the
ridgel. 

\begin{figure} 
  \includegraphics[width=6.0cm]{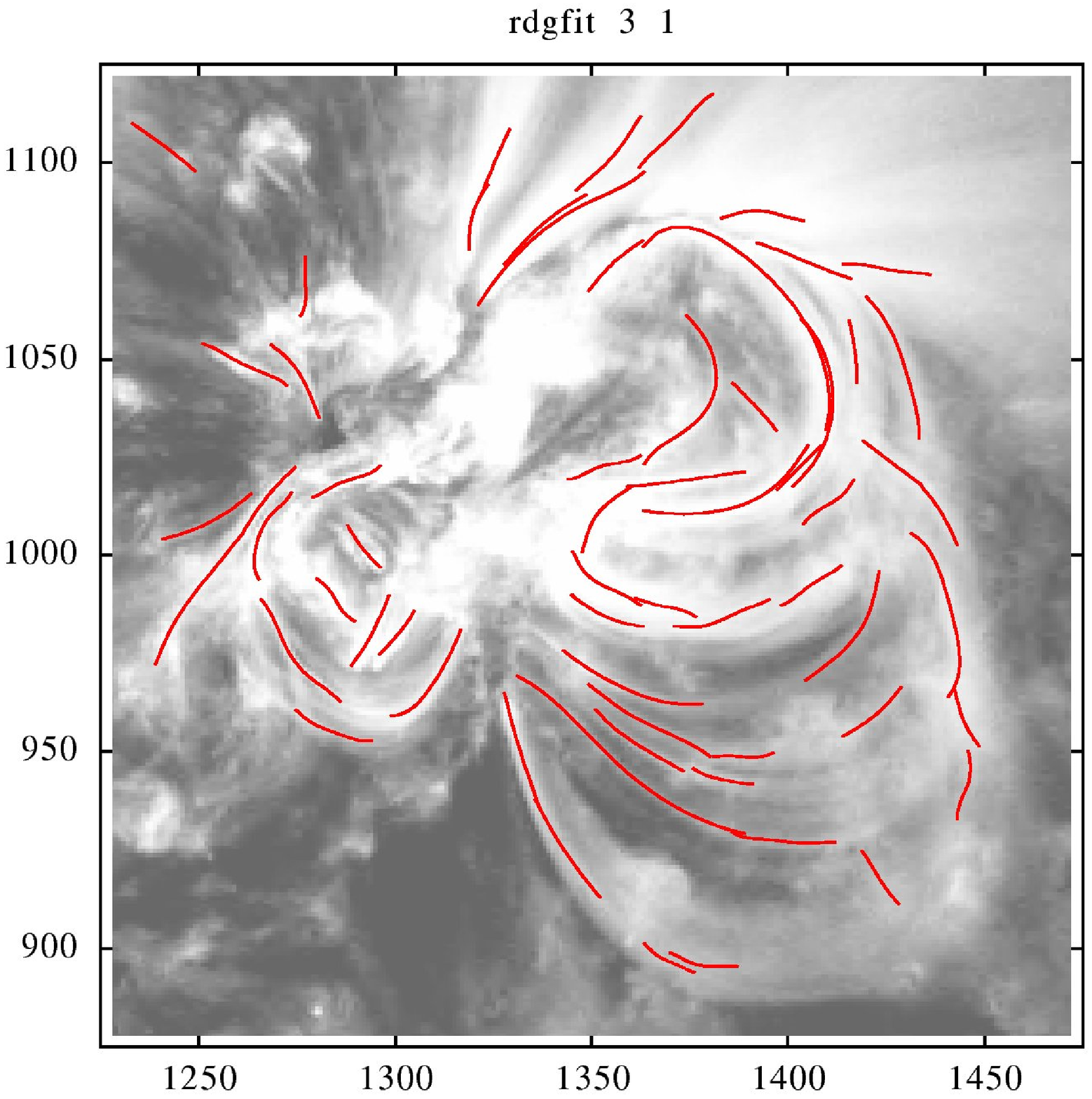}
  \includegraphics[width=6.0cm]{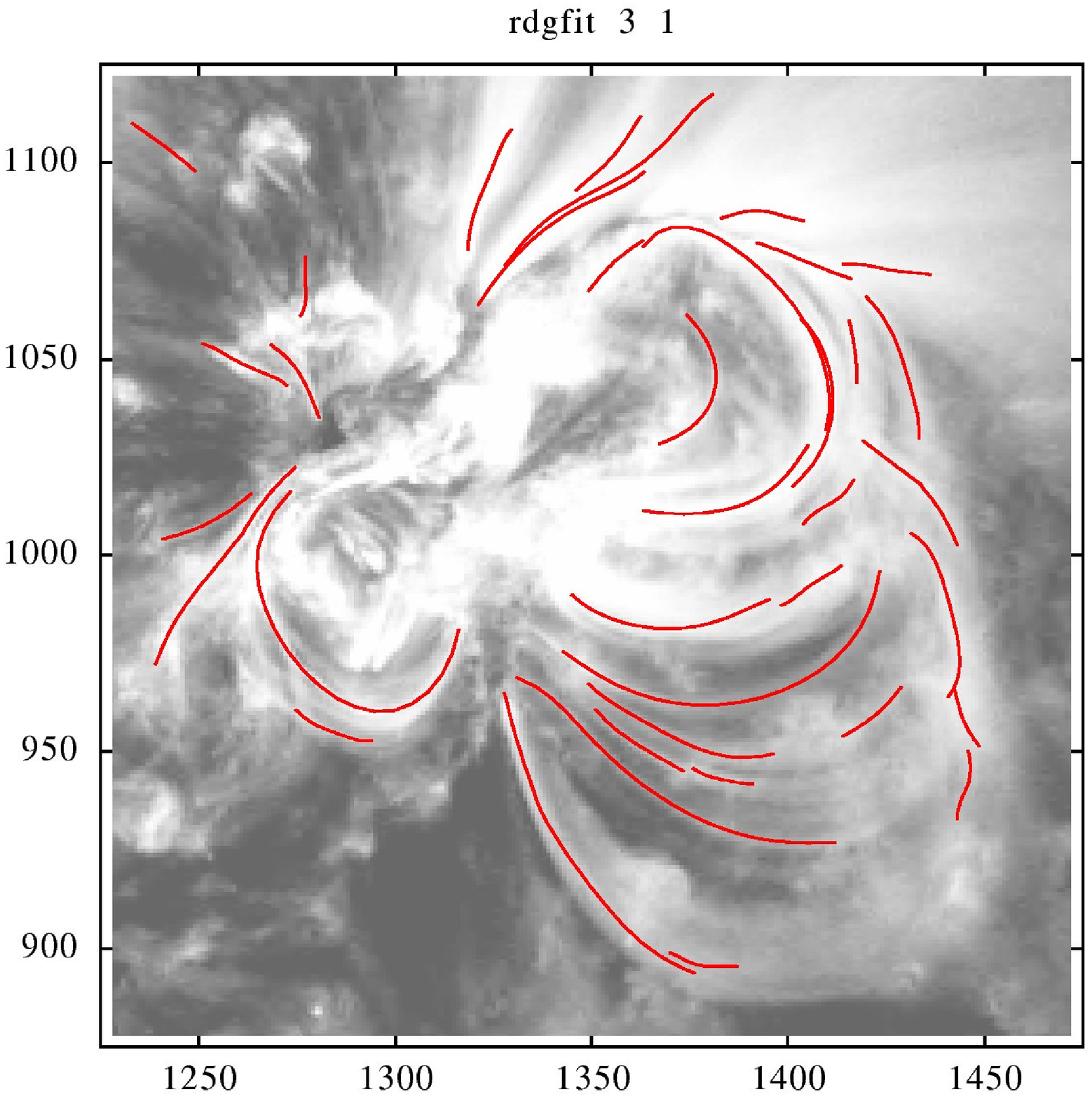}
\caption{Fit curves obtained for those chains which involve ten or more
  ridgels (left) and curves remaining after cleaning of those curves
  which are due to moss (right).}\label{fig:fits}
\end{figure}

For different $\mu$ this minimum yields fit curves with different
levels of curvature. The local inverse radius of curvature can at any point
along the curve be calculated from (\ref{eq:polynomial}) by
\begin{equation}
  \frac{1}{r(t)}=\frac{|\vect{p''}(t)\times\vect{p'}(t)|}{|\vect{p'}(t)|^3}
\end{equation}
The final $\mu$ is then chosen so that
\begin{equation}
     E_\mathrm{chain}(\mu) = \sum_{i\in\mathrm{chain}} \frac{d_i^2}{d_\mathrm{max}^2}
  \;+\;\sum_{i\in\mathrm{chain}} \frac{\Delta\alpha_i^2}{\alpha_\mathrm{max}^2}
  \;+\;r_\mathrm{min}^2 \int\limits_{-1}^1 \frac{1}{r(t)^2}\;dt
\label{eq:fitenergy}\end{equation}
is a minimum where $\Delta\alpha_n$ is the angle between the local
normal direction of $\vect{d}_n$ of the fit curve and the ridgel
orientation $\pm\vect{u}_{\perp,n}$, the sign again chosen to yield
the smallest possible $\abs{\Delta\alpha_i}$
The meaning of the terms is obvious, and clearly the first two terms
in general require a large curvature which is limited by the minimization
of the last term.

Expression (\ref{eq:fitenergy}) depends nonlinearly on the parameter
$\mu$ which we use to control the overall curvature. The minmum for
(\ref{eq:fitenergy}) is found by iterating $\mu$ starting from a
large numerical value, {\it i.e.}, a straight line fit. The parameters
$r_\mathrm{min}$, $\alpha_\mathrm{max}$, and $d_\mathrm{max}$ can be
used to obtain fits with a different balance between the mean square
spatial and angular deviation of the fit form the ``observed'' chain
of ridgels and the curvature of the fit. Unless these parameters are
chosen reasonably, {\it e.g.} $r_\mathrm{min}$ not too small, we have
always found a minimum for (\ref{eq:fitenergy}) after a few iteration
steps.

In the left part of Figure~\ref{fig:fits} we show the final fits
obtained. For this result, the ridgels were found by automated
scaling and interpolated by method (\ref{eq:ridgept1}), the parameters
in (\ref{eq:cocirc}) and (\ref{eq:fitenergy}) were $h_\mathrm{max}=
3.0$ pixels, $r_\mathrm{min}=15.0$ pixels, and $a_\mathrm{max}=10.0$
degrees. The fits are represented by fifth degree parametric
polynomials.

Obviously, the image processing cannot easily distinguish between
structures which are due to moss and bright surface features and
coronal loops. Even the observer is sometimes misled and there are no
rigorous criteria for this distinction. Roughly, coronal loops produce
longer and smoother fit curves, but there is no strict threshold because
it may appear that the fit curve is split along a loop where the loop signal
becomes faint.
As a rule of thumb, a restriction to smaller curvature by choosing a
higher parameter $r_\mathrm{max}$ and discarding shorter fit curves
tends to favour coronal loops. Eventually, however, loops are
suppressed, too.
We have therefore appended a user interactive tool as the last step of
our processing which allows us to eliminate unwanted curves, merge or split
curves when smooth fits result with an energy (\ref{eq:fitenergy}) of
the output fits not much higher than the energy of the input.
The left part of Figure~\ref{fig:fits} shows the result of such a
cleaning step.

\section{Application}
\label{Application}

\begin{figure}
 \hspace*{\fill}
  \includegraphics[bb=30 40 556 450,clip,width=9cm]{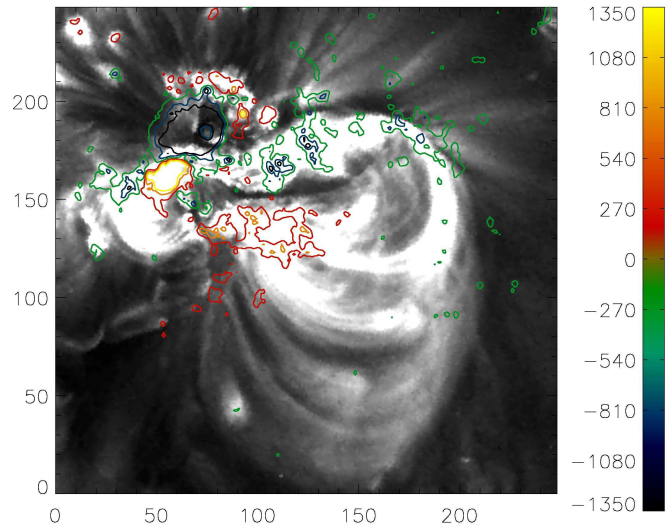}
 \hspace*{\fill}
\caption{MDI contours overlaid on the STEREO/SECCHI EUV image for NOAA 10930.
  The EUV and MDI data were recorded on 12 December 2006 at 23:43:11 UT and
  23:43:30 UT, respectively. The colour code on the right indicates the field
  strength at the coutour lines in Gauss.}
\label{mdi_euvi}
\end{figure}

In this section we present an application of our segmentation tool to
another EUV image of active region NOAA 10930 taken by the SECCHI instrument
onboard STEREO A.  This EUV image was observed at
$\lambda$ = 195 {\AA} on 12 December 2006 at 23:43:11 UT. 
At that time the STEREO spacecraft were still close so that stereoscopy
could not be applied. We therefore selected an image which was taken
close to the MDI magnetogram observed at 23:43:30 UT on the same day.
It is therefore possible to calculate magnetic-field lines from an
extrapolation model and project them onto the STEREO view direction
to compare them with the loop fits obtained with our tool. 
In Figure~\ref{mdi_euvi} the MDI contour lines of the line-of-sight field
intensity were superposed on the EUV image.

\begin{figure}
  \includegraphics[width=6cm]{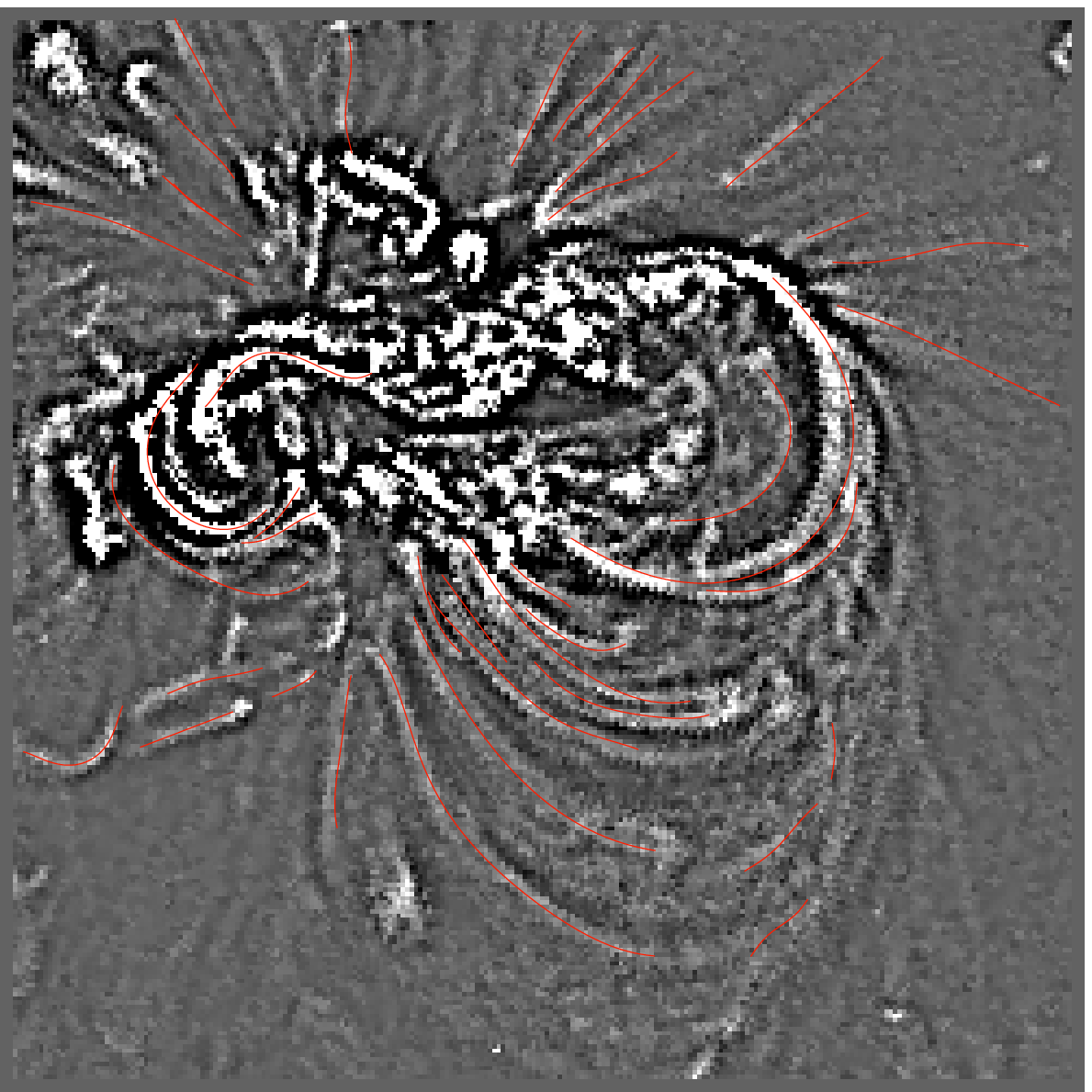}
  \includegraphics[width=6cm]{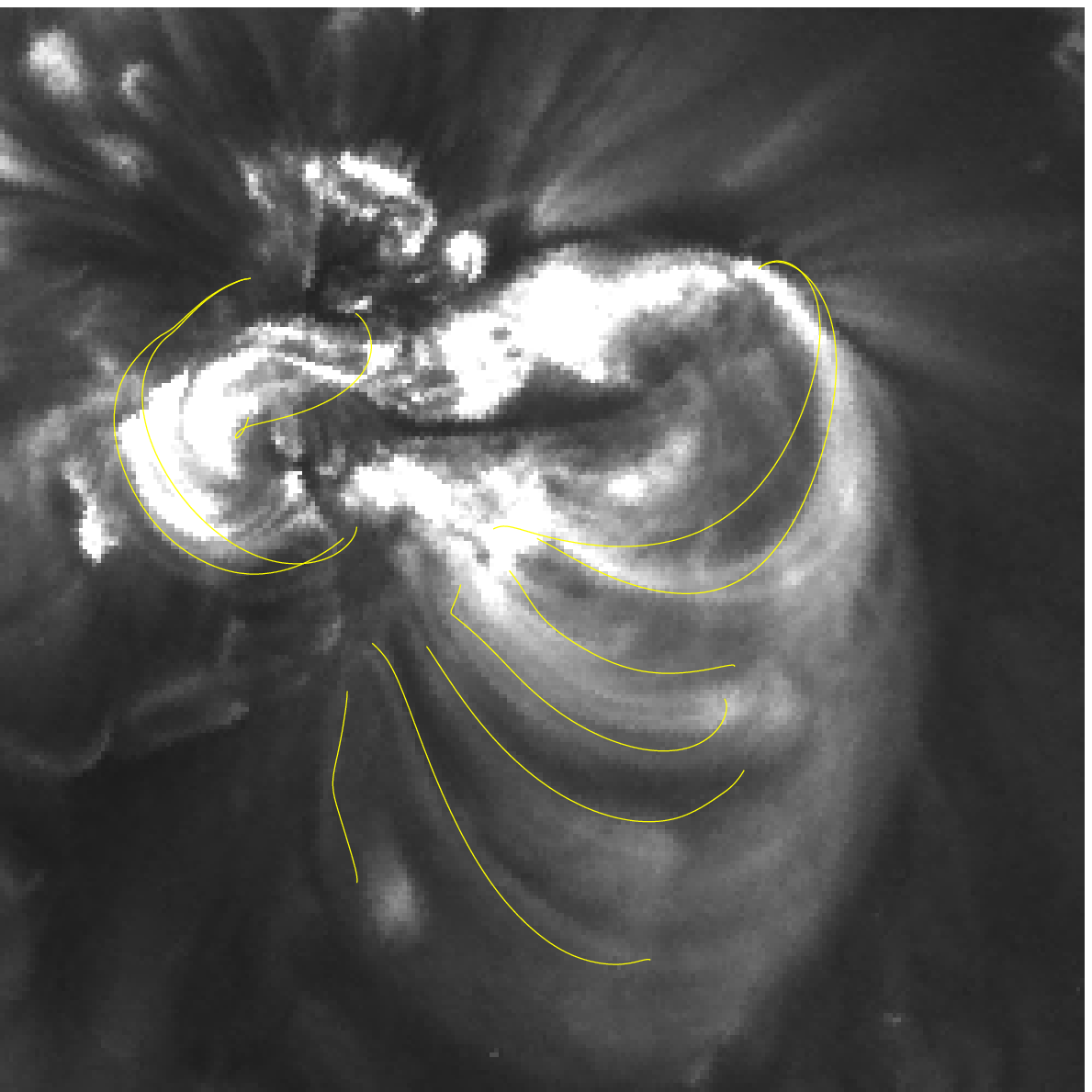}
\caption{The left diagram shows as red lines the loops identified
  by the segmentation tool for the EUV image in Figure~\ref{mdi_euvi}.
  In order to show the loops more clearly, the image in the background
  was contrast enhanced by an unsharp mask filter.
  The right diagram displays in yellow field lines calculated from the
  MDI data.
  The fieldlines were selected so that they are located closest to the
  extracted loops in the left part of the image. See the text for more
  details on how the field lines were determined.}
\label{line_euvi}
\end{figure}

\begin{figure} 
  \includegraphics[bb=40 25 328 328,clip,width=6cm]{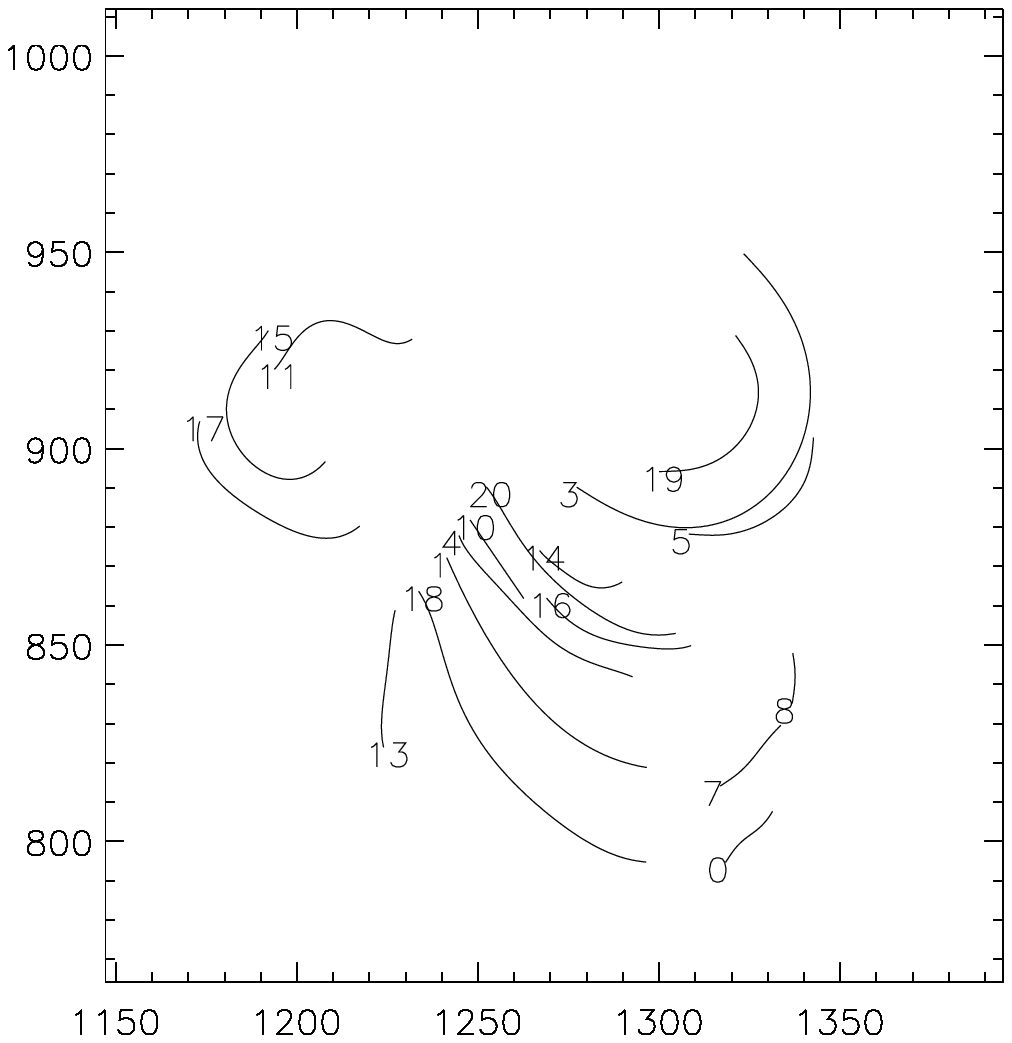}
  \includegraphics[bb=40 25 328 328,clip,width=6cm]{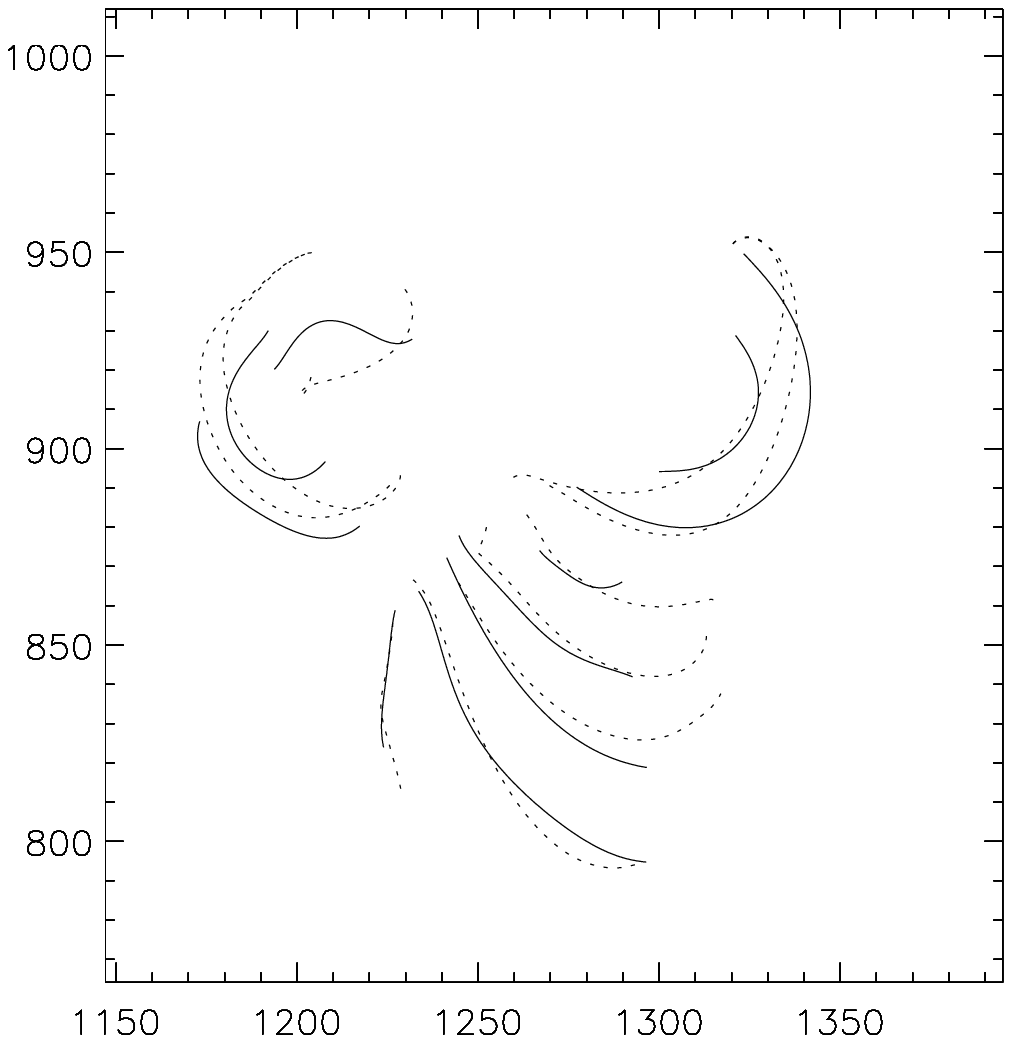}
\caption{The left panel shows the loops identified by the segmentation tool
which corresponds to the closed magnetic filed lines; In the right panel
are the loops (solid lines) with their best-fitting field lines(dotted lines).
$x$ and $y$ axis are in units of EUV pixels.}
\label{loop_line}
\end{figure}

The loop fits here were obtained by applying the automated scaling
with $d$ up to two pixels, {\it i.e.} window sizes up to $2d+1=5$ pixels,
to identify the ridgels. Pixels with maximum quality $q$ below 0.4
were discarded and we applied method (\ref{eq:ridgept1})
to interpolate the local ridge maxima.
$h_\mathrm{max}=5.0$ pixels,
$r_\mathrm{min}=15.0$ pixels, and
$a_\mathrm{max}=10.0$ degrees.
The fits are fifth degree parametric polynomials.
In Figure~\ref{line_euvi}, we show some of the fits obtained which are most
likely associated with a coronal loop. They are superposed onto the
EUV image as red lines. Those loops which were found close to computed
magnetic-field lines are displayed again in the left part of
Figure~\ref{loop_line} with loop numbers so that they can be identified.

The magnetic-field lines were computed from the MDI data by an
extrapolation based on a linear force-free field model
(see \opencite{seehafer78} and \opencite{alissandrakis81} for details).
This model is a simplification of the general nonlinear force-free
magnetic-field model
\begin{equation}
\nabla \times {\bf B } = \alpha {\bf  B}
\quad\text{where}\quad
{\bf B} \cdot \nabla \alpha  = 0
\nonumber\end{equation}
and $\alpha$ may vary on different field lines. An extrapolation
of magnetic surface observations based on this model requires boundary
data from a vector magnetograph.
The linear force-free field model treats $\alpha$ as a global constant.
The advantage of linear the force-free field model is that it requires only
a line-of-sight magnetogram, such as MDI data, as input.

A test of the validity of the linear forec-free assumption is to
determine different values of $\alpha$ from a comparison of field
lines with individual observed loop structures
({\it e.g.}, \opencite{carcedo:etal03}). The range of $\alpha$ obtained then
indicates how close the magnetic field can be described by the linear
model. Since the linear force-free field has the minimum energy for
given normal magnetic boundary field and magnetic helicity, a linear
force-free field is supposed to be much more stable than the more
general non-linear field configuration \cite{taylor74}.

We calculated about 5000 field lines with the linear force-free model
with the $\alpha$ value varied in the range from -0.0427 Mm$^{-1}$ to
0.0356 Mm$^{-1}$. These field lines were then projected onto the EUV
image for a comparison with the detected loops.


For each coronal loop $l_{i}$, we calculate the average distance of
the loop to every projected field line $b_{j}$
discarding those field lines that do not fully cover the observed
loop $l_{i}$. This distance is denoted by $C_{l_{i}}(b_{j})$. 
For details of this distance calculation see \inlinecite{feng:etal:2007}. In
the end we could find a closest field line for every coronal loop
by minimizing $C_{l_{i}}(b_{j})$. The
detected loops and their closest field lines are plotted in the right
diagram of Figure~\ref{loop_line}.
An overplot of the closest field lines onto the EUV image is shown in
Figure~\ref{line_euvi}

In Table~\ref{match-alpha} we list the distance measure $C$ along
with the loop number and the linear force-free parameter $\alpha$
for the closest field line found. 
We find that our $\alpha$ values are not uniform
over this active region, that is, the linear force-free model is not
adequate to describe the magnetic properties of this active region.
This is also seen by the characteristic deviation at their the upper
right end in Figure~\ref{line_euvi} between the eastwards-inclined loops
(solid) and their closest, projected, field lines (dotted). With no value
of $\alpha$ the shape of these loops could be satisfactorily fitted.
Further evidence for strong and inhomogeneous currents in the
active region loops is provided by the fact that only 2.5 hours later,
at 02:14 UT on 13 December, a flare occurred in this active region
and involved the magnetic structures associated with loops 2, 4, and 17. 

\begin{table}
\begin{tabular}{rrr} \hline
Loop No. & $C_{l}({\rm b})$(pixel) & $\alpha$($10^{-3}Mm^{-1}$) \\
\hline
1    &3.1957  &-10.680  \\
3    &3.7523  &-35.600  \\
4    &2.3432  &-9.2560  \\
11  &10.7692  &-35.600  \\
13   &0.4864  &2.1360  \\
14   &1.3636  &-9.2560  \\
15   &4.2386  &17.088   \\
17   &4.8912  &16.376   \\
18   &2.4256  &-14.240  \\
19   &2.5388  &-32.752  \\
\hline
\end{tabular}
\caption{Identified loops, $\alpha$ values of the best fitting field lines and
the averaged distances in units of pixel between the loop and the best fitting
field line. }
\label{match-alpha}
\end{table}

\section{Discussion}

EUV images display a wealth of structures and there is an important need
to reduce this information for specified analyses. For the study of
the coronal magnetic field, the extraction of loops from EUV images is
a particularly important task. Our tool intends to improve earlier
work in this direction.
Whether we have achieved this goal can only be decided from a rigourous
comparison which is underway elsewhere \cite{aschw:etal:2007}. At least
from a methodological point of view, we expect that our tool should
yield improved results compared to Strous (2002, unpublished) and
\inlinecite{Lee:etal:2006}.

From the EUV image alone it is often difficult to decide which of the
features are associated with coronal loops and which are due to moss
or other bright surface structures. A final comparison of the loops
with the extrapolated magnetic field and its field line shapes is
therefore very helpful for this distinction. Yet we have avoided to
involve the magnetic-field information in the segmentation procedure
which extracts the the loops from the EUV image because this might
bias the loop shapes obtained.

For the case we have investigated,
we find a notable variation of the optimal $\alpha$ values and
also characteristic deviation of the loop shapes from the calculated
field lines. 
These differences are evidence of the fact that the true coronal magnetic
field near this active region is not close to a linear force-free state.
This is in agreement with earlier findings. \inlinecite{wiegelmann:etal05}, {\it e.g.},
have shown for another active region that a nonlinear force-free model
describes the coronal magnetic field more accurately than linear models.
The computation of nonlinear models is, however, more involved due to
the nonlinearity of the mathematical equations
({\it e.g.}, \opencite{wiegelmann04}; \opencite{inhester:etal:2006}).
Furthermore, these models require photospheric vector magnetograms as
input, which were not available for the active region investigated.

Coronal loops systems are often very complex. In order to acess them
in 3D, the new STEREO/SECCHI telescopes now provides EUV images which
can be analysed with stereoscopic tools. We plan to apply our
loop-extraction program to EUV images from different viewpoints and
undertake a stereoscopic reconstruction of the true 3D structure of
coronal loops along the lines described by \inlinecite{inhester:2006} and
\inlinecite{feng:etal:2007}.
The knowledge of the 3D geometry of a loop allows to estimate more precisely
its local EUV emissivity. From this quantity we hope to be able to derive
more reliably the plasma parameters along the length of the loop.

Other applications can be envisaged. An interesting application of our
tool, {\it e.g.}, will be the investigation of loop oscillations. Here, the
segmentation tool will be applied to times series of EUV images. We
are confident that oscillation modes and in the case of a STEREO/SECCHI
pairwise image sequence the polarisation of the loop oscillation
can also be discerned.

\begin{acks}
BI thanks the International Space Institute, Bern, for their hospitality
and the head of its STEREO working group, Thierry Dudoc de Wit and
also Jean-Francois Hochedez for stimulating discussions.
LF was supported by the IMPRESS graduate school run jointly
by the Max Planck Society and the Universities G\"ottingen and Braunschweig.
The work was further supported by DLR grant 50OC0501.

The authors thank the SOHO/MDI and the STEREO/SECCHI consortia for
their data. SOHO and STEREO are a joint projects of ESA and
NASA. The STEREO/ SECCHI data used here were produced by an
international consortium of the Naval Research Laboratory (USA),
Lockheed Martin Solar and Astrophysics Lab (USA), NASA Goddard Space
Flight Center (USA) ,Rutherford Appleton Laboratory (UK), University of
Birmingham (UK), Max-Planck-Institut for Solar System Research(Germany),
Centre Spatiale de Liege (Belgium), Institut d'Optique Théorique et
Appliqueé (France), Institut d'Astrophysique Spatiale (France).

The USA institutions were funded by NASA; the UK institutions by
Particle Physics and Astronomy Research Council (PPARC); the German
institutions by Deutsches Zentrum f\"ur Luft- und Raumfahrt e.V. (DLR);
the Belgian institutions by Belgian Science Policy Office; the French
institutions by Centre National d~REtudes Spatiales (CNES) and the
Centre National de la Recherche Scientifique (CNRS). The NRL effort
was also supported by the USAF Space Test Program and the Office of
Naval Research.

\end{acks}


\end{article}
\end{document}